\documentclass[a4paper,12pt]{article}
\usepackage{graphicx}
\usepackage{amsmath}
\usepackage[cp1251]{inputenc}
\usepackage[russian,english]{babel}

\begin{document}
\selectlanguage{russian}
\begin{otherlanguage}{english}
\begin{center}
{\bf Model for synchronizer of marked pairs in fork-join network}
\\
\bigskip

S.~V.~Vyshenski\dag\footnote[1]{svysh@pn.sinp.msu.ru},
P.V.~Grigoriev\ddag, and
Yu.Yu.~Dubenskaya\S
\\
\bigskip

\dag~Institute of Nuclear Physics, Moscow State University, Moscow 119899,
Russia \\
\ddag~General Physics Institute, Russian Academy of Sciences, Vavilov str., 38,
Moscow 119991, Russia\\
\S~Institute of Precise Mechanics and Computer Technology, Russian Academy of Sciences,
Leninskiy av., 51, Moscow 119991, Russia
\end{center}

\bigskip

\begin{abstract}
We introduce a model for synchronizer of marked pairs, which is a node
for joining results of parallel processing in
two-branch fork-join queueing network.
A distribution for number of jobs in the synchronizer is obtained.
Calculations are performed assuming that:
arrivals to the network form a Poisson process,
each branch operates like an M/M/N queueing system.
It is shown that a mean quantity of jobs in the synchronizer
is bounded below by the value, defined by
parameters of the network (which contains the synchronizer)
and does not depend upon performance and particular properties
of the synchronizer.
A domain of network parameters is found, where
the flow of jobs departing from the synchronizer
does not manifest a
statistically significant difference from the Poisson type,
despite the
correlation between job flows from both branches of the fork-join network.

\end{abstract}

\end{otherlanguage}

\bigskip
\pagebreak

\begin{center}
{\bf Модель синхронизатора маркированных пар
в сети разветвление-объединение
}
\bigskip

С.В.~Вышенский\dag$^1$,
П.В.~Григорьев\ddag,
Ю.Ю.~Дубенская\S
\\
\bigskip

\dag~НИИ ядерной физики МГУ, Москва 119899\\
\ddag~Институт общей физики РАН, Вавилова 38, Москва 119991\\
\S~Институт точной механики и вычислительной техники РАН, Ленинский просп. 51, Москва 119991
\\
\bigskip


Статья публикуется в журнале\\
\emph{Обозрение прикладной и промышленной математики, 2008}

\end{center}

\begin{abstract}
Предложена модель синхронизатора маркированных пар --
узла объединения результатов параллельной обработки двух потоков в сетях
массового обслуживания типа разветвление-объединение (fork-join).
Найдено распределение количества заявок в синхронизаторе.
Расчеты проведены для стационарного режима в следующих предположениях:
на вход сети поступает поток заявок пуассоновского типа,
системы в обеих ветвях сети относятся к типу $M/M/N$.
Показано, что среднее число заявок в
синхронизаторе ограничено снизу значением, которое определяется параметрами сети,
содержащей синхронизатор,
и не зависит от производительности и особенностей синхронизатора.
Найдена область параметров сети,
в которой корреляция между потоками заявок из разных ветвей сети не приводит к статистически
значимому отклонению потока на выходе синхронизатора от пуассоновского типа.
\end{abstract}

\tableofcontents
\pagebreak

\section{Введение}

В работе \cite{SynchroIdeal2007} предложен новый функциональный элемент
\emph{синхронизатор маркированных пар} для описания процесса
объединения результатов параллельной обработки двух потоков заявок в сетях
массового обслуживания типа разветвление-объединение (РО) или fork-join.
Оказалось, что даже самое абстрактное представление о синхронизаторе
позволяет применить известный математический аппарат для анализа
нагрузочных характеристик сети РО, которая содержит такой синхронизатор.
В настоящей работе предложена модель, в минимальной степени
детализирующая устройство синхронизатора. Введение такой модели
позволяет исследовать дополнительные свойства сети РО. В частности --
провести анализ непуассоновских свойств потоков заявок в сетях РО.

\begin{figure}
\begin{center}
\includegraphics[width=10cm]{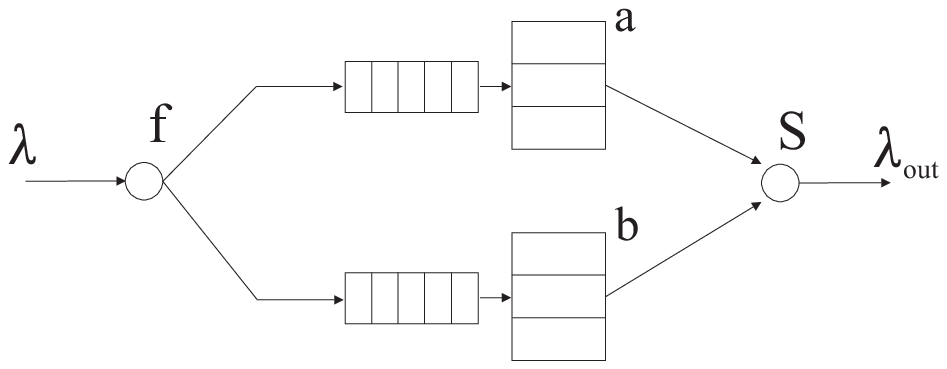}
\end{center}
\caption{
Сеть разветвление-объединение с двумя ветвями.}
\label{PictForkJoinDouble}
\end{figure}

Рассмотрим сеть
массового обслуживания типа РО,
показанную на рисунке \ref{PictForkJoinDouble}.
Предположим, что на вход сети поступает пуассоновский поток промаркированных
(например, пронумерованных)
заявок с интенсивностью $\lambda > 0$.
В точке разветвления $f$ каждая заявка разделяется на две заявки с одинаковыми номерами,
совпадающими с номером исходной заявки.
Эти две заявки одновременно поступают на вход ветвей $a$ и $b$,
которые представляют собой системы массового обслуживания
$M/M/N_a$ и $M/M/N_b$, где $N_a, \,N_b \geq 1$ задают количества
параллельных каналов обслуживания в ветвях $a$ и $b$.
Очереди в ветвях $a$ и $b$ подчиняются дисциплине FIFO.
Сеть функционирует в стационарном режиме.
Для объединения результатов параллельной обработки двух потоков в
ветвях $a$ и $b$ сети служит узел $S$, названный в нашей работе \cite{SynchroIdeal2007}
\emph{синхронизатором маркированных пар}.

Сеть массового обслуживания РО представляет собой удобную модель для расчета
нагрузочных характеристик фрагментов различных информационных,
коммуникационных и производственных систем.
В частности, такой фрагмент характерен для систем, функциональность которых
принято задавать на языке потоков работ (workflow).

В большинстве работ о сетях РО
\cite{Flatto1984,
Baccelli1985,
Raghavan2001,
Nelson1988,
Varma1994,
Ayhan2001,
Knessl1991,
Nguyen1993,
Serfozo2004}
решается задача вычисления
времени отклика сети, то есть промежутка времени, разделяющего моменты разветвления
и объединения потоков заявок в ветвях сети,
а также распределение для общего количества заявок в сети РО.
В этих работах не конкретизировался механизм объединения потоков заявок и
не исследовались статистические свойства этого процесса.

В настоящей работе предлагается
модель для синхронизатора маркированных пар
как отдельного функционального элемента сети РО.
К основным результатам настоящей статьи относятся:
представление идеального синхронизатора маркированных пар
в виде системы массового обслуживания с бесконечным количеством
каналов обслуживания,
нахождение области параметров сети, в которой отличие выходного потока
синхронизатора от пуассоновского не является статистически значимым.
В той же области параметров предложено приближение для распределения количества заявок
в синхронизаторе.
Полученные результаты позволяют решать следующие задачи:
расчет ресурсов для бесперебойной работы сети РО,
оптимизация ресурсов для работы сети РО,
исследование взаимодействие сети РО с
более крупной сетью, содержащей данную сеть РО.

Дальнейшее изложение построено следующим образом.
В разделе \ref{SectModels} приведен общий алгоритм работы синхронизатора и
предложена модель идеального (бесконечно быстрого) синхронизатора маркированных пар.
В разделе \ref{SectIn} исследуются статистические свойства потоков заявок
на входе синхронизатора.
Для сети РО при $N_a = N_b = 1$
путем численного решения уравнений
Колмогорова-Чепмена показано, что поток на входе синхронизатора не является
пуассоновским.
Для сети РО с параметрами $N_a, \,N_b > 1$
методам моделирования показано, что входной поток
синхронизатора является \textit{почти пуассоновским} (см. Приложение \ref{SectStat})
при некоторых комбинациях параметров сети.
В разделе \ref{SectQuant} почти пуассоновское приближение
для входного потока синхронизатора
позволило применить известный математический аппарат для описания
распределения количества
требований в синхронизаторе. Показано, что
даже в идеальном синхронизаторе
среднее число заявок в синхронизаторе ограничено снизу значением,
которое определяется лишь параметрами сети РО, содержащей синхронизатор,
и не зависит от свойств синхронизатора.
В разделе \ref{SectOut}
найдены условия,
при которых поток на выходе синхронизатора является почти пуассоновским.
В разделе \ref{SectDiscus} полученные в настоящей статье результаты
сравниваются с известными из литературы, а также обсуждаются возможные
применения.
В Приложении \ref{SectStat} дано определение почти пуассоновского потока, а также
изложены использованные методы численного моделирования
для проверки статистических гипотез.

\section{Модель синхронизатора маркированных пар} \label{SectModels}

Процесс объединения двух потоков маркированных заявок в разных сетях может происходить
по-разному. В настоящей работе мы предлагаем приближенное описание этого процесса
с помощью нового функционального элемента -- синхронизатора
$S$ маркированных пар (рис. \ref{PictForkJoinDouble}).
Будем считать, что этот синхронизатор $S$ состоит из
памяти синхронизатора и монитора $M$,
который отслеживает и изменяет состояние памяти синхронизатора.
Полагаем, что монитор $M$ срабатывает мгновенно,
а размер памяти неограничен.
Именно поэтому мы называем синхронизатор \emph{идеальным}.
Идеальный синхронизатор работает следующим образом.

Заявки из ветвей $a$ и $b$ поступают на вход синхронизатора $S$ маркированных пар
и сохраняются в его памяти.
Пару заявок различных типов ($a$ и $b$) с одинаковыми номерами будем
называть \emph{партнерами}.
Партнера, достигшего синхронизатора первым из пары партнеров, будем называть \emph{первым партнером}.
Партнера, достигшего синхронизатора вторым из пары партнеров, будем называть \emph{вторым партнером}.

После сохранения вновь поступившей заявки в памяти, монитор $M$ производит
следующие действия:

\begin{itemize}
\item определяется номер и тип ($a$ или $b$) новой заявки,
\item если в памяти синхронизатора уже имеется (первый) партнер для новой заявки,
то оба партнера найденной пары удаляются из памяти
синхронизатора и передаются на выход синхронизатора.
\item если партнер для новой заявки в памяти отсутствует, то новая заявка
(то есть первый партнер из пары)
остается в памяти синхронизатора ждать своего второго партнера.
\end{itemize}

Исходя из приведенного алгоритма работы синхронизатора, определим
время $t$ пребывания пары в синхронизаторе как разницу во
времени между приходом второго и первого партнеров из пары.

Функционирование систем $a$ и $b$ описывается с помощью классической теории массового обслуживания
и не представляет сложности.
Устройство идеального синхронизатора можно описать при помощи следующей модели.

\subparagraph*{Модель $S_0$ синхронизатора как системы массового обслуживания
с бесконечным числом каналов.}

\begin{figure}
\begin{center}
\includegraphics[width=8cm]{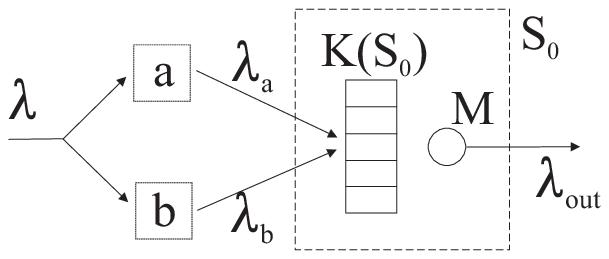}
\end{center}
\caption{
Модель $S_0$ синхронизатора как системы массового обслуживания с бесконечным числом каналов.
}
\label{PictSyncQS}
\end{figure}

На рисунке \ref{PictSyncQS}
память синхронизатора реализована в виде бесконечного числа параллельных
каналов обслуживания. $K(S_0)$ -- число каналов, занятых в некоторый момент времени.

В этой модели \emph{временем обслуживания} является время $t$ пребывания в синхронизаторе.
В течение этого времени первый партнер ожидает
в синхронизаторе
своего второго партнера.
Заметим, что первый партнер покидает синхронизатор, как только приходит его
второй партнер, вне зависимости от количества остальных первых
партнеров в синхронизаторе и их времени ожидания. Тем самым, первые партнеры
не образуют очереди в традиционном смысле этого термина,
а моменты выхода из синхронизатора определяются только временем
обслуживания.

Таким образом, в терминах модели $S_0$ синхронизатор представлен в виде
``виртуальной'' системы
массового обслуживания с бесконечным числом каналов. Виртуальность означает, что
реального обслуживания заявок не производится. Просто первые партнеры ожидают вторых
партнеров.
При этом поступление в синхронизатор первого партнера определяет момент начала обслуживания пары,
а поступление в синхронизатор второго партнера определяет момент окончания обслуживания пары.
Можно сказать, что \emph{виртуальный} поток in$(S_0)$ заявок на входе модели $S_0$ совпадает с реальным потоком первых партнеров пар, поступающих в синхронизатор.
А \emph{виртуальный} поток заявок out$(S_0)$ на выходе модели $S_0$ совпадает с
реальным потоком вторых партнеров пар, поступающих в синхронизатор.

Такая модель позволит нам применить для расчета характеристик синхронизатора
хорошо известный математический аппарат \cite{Kleinrock1979},
развитый для систем $G_1/G_0/\infty$, где
$G_1$ -- произвольный закон распределения интервалов между приходами первых партнеров,
$G_0$ -- произвольный закон распределения времени $t$ пребывания в синхронизаторе.

\section{Поток на входе синхронизатора} \label{SectIn}

Поток заявок на входе модели $S_0$
(то есть виртуальный поток in$(S_0)$)
совпадает с потоком
первых партнеров маркированных пар.
В стационарном режиме сети РО
скорость $\lambda_1$ поступления в синхронизатор первых партнеров каждой пары:
$
\lambda_1=p_a \lambda_a + p_b \lambda_b = \lambda
$,
где $\lambda_a =\lambda_b=\lambda$ -- скорость поступления заявок
в ветвь $a$ или $b$, а $p_a$ и $p_b=1-p_a$ -- вероятности прихода
первого партнера в синхронизатор из ветви $a$ или $b$ соответственно.
Таким образом,
интенсивность потока in$(S_0)$ равна
$\lambda_{\text{in}} = \lambda$.

Если бы поток in$(S_0)$
оказался пуассоновским, то это позволило бы нам
заменить модель $S_0 = G_1/G_0/\infty$ более простой моделью $S_0 = M/G_0/\infty$, а значит,
явно рассчитать распределение количества заявок в синхронизаторе \cite{Adan2002}.

Покажем, однако, что поток первых партнеров не всегда является пуассоновским.
Рассмотрим сеть РО с двумя ветвями $\{M/M/1$; \, $M/M/1\}$
в каждой из которых имеется по одному
каналу обслуживания $N_a = N_b = 1$, а
на вход сети поступает пуассоновский поток интенсивности $\lambda$
(рис. \ref{PictForkJoinDouble}).
Как мы ожидаем, при $N_a = N_b = 1$ отклонения потока in$(S_0)$
от пуассоновского максимальны.
В стационарном режиме
состояние ветви $i$ ($i$ принимает значения $a$ или $b$) в некоторый момент времени
будем характеризовать неотрицательным целым числом $q_i$ --
количеством требований, находящихся в ветви $i$.
Состояние обеих ветвей, рассматриваемых совместно,
характеризуется парой чисел $(q_a,\, q_b)$. Вероятность
состояния $(q_a,\, q_b)$ равна $P(q_a,\, q_b)$.
Интенсивности
обслуживания в ветвях равны $\mu_{a}$ и $\mu_{b}$.
Обозначим $\psi_{a}=\lambda/(N_a \mu_a)$ и $\psi_{b}=\lambda/(N_a \mu_b)$.
В стационарном режиме
вероятности $P(q_a,\, q_b)$ не зависят от времени, а
$\psi_{a}<1$ и $\psi_{b}<1$.

Найдем распределение вероятностей
$P(q_{a},q_{b})$ для произвольного
фиксированного набора параметров сети РО:
$\lambda,\, \mu_a,\, \mu_b$.
Эта задача решалась численно \cite{SynchroIdeal2007} путем итерационного поиска стационарного решения
системы уравнений Колмогорова-Чепмена для сети РО $\{M/M/1$; \, $M/M/1\}$:

\noindent
при $q_{a}=q_{b}=0$:\qquad $\lambda P(0,0)=\mu_{a}P(1,0)+\mu_{b}P(0,1)$,\\
при $q_{b}>0$:\qquad
$(\lambda+\mu_{b})P(0,q_{b})=\mu_{a}P(1,q_{b})+\mu_{b}P(0,q_{b}+1)$,\\
при $q_{a}>0$:\qquad
$(\lambda+\mu_{a})P(q_{a},0)=\mu_{a}P(q_{a}+1,0)+\mu_{b}P(q_{a},1)$,\\
при $q_{a} q_{b}>0$:\qquad
$(\lambda+\mu_{a}+\mu_{b})P(q_{a},q_{b})=\lambda P(q_{a}-1,\,q_{b}-1)+\mu_{a}P(q_{a}+1,\,q_{b})+\mu_{b}P(q_{a},\,q_{b}+1)$.

Используя решения этих уравнений,
покажем, что поток in$(S_0)$
на входе синхронизатора отличается от пуассоновского потока.

Рассмотрим
условную вероятность $P_{\text{cond}}$ появления первого партнера в течение
малого интервала времени
$d\tau$ сразу после появления предыдущего первого партнера:
$$P_{\text{cond}}=\frac{
        \mu_{a}^2P(q_{b}>q_{a}>1) +
        \mu_{b}^2P(q_{a}>q_{b}>1) +
        (\mu_{a}^2+\mu_{b}^2)P(q_{a}=q_{b}>1)
    }
    {
        \mu_{a}P(q_{b}>q_{a}>0) +
        \mu_{b}P(q_{a}>q_{b}>0) +
        (\mu_{a}+\mu_{b})P(q_{a}=q_{b}>0)
    } d\tau.
$$

\begin{figure}
\begin{center}
\includegraphics[width=12cm]{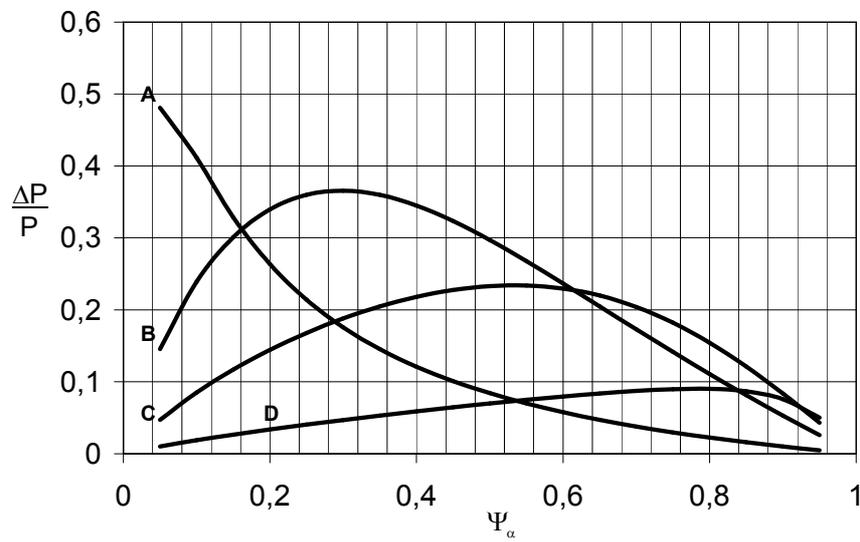}
\end{center}
\caption{Отклонение $\Delta P/P$ потока in$(S_0)$ от пуассоновского
для сети $N_a = N_b = 1$ в зависимости
от $\psi_{a}$ при $\psi_{b}=0,05$ (A),
$\psi_{b}=0,35$ (B), $\psi_{b}=0,65$ (C)
и $\psi_{b}=0,90$ (D).
}
\label{PictPuass}
\end{figure}

Для пуассоновского потока с интенсивностью $\lambda$,
вероятность $P_{\text{cond}}$
не отличалась бы от безусловной вероятности $P=\lambda d\tau$
появления события в течение интервала времени $d\tau$.
Однако, как показывает решение уравнений Колмогорова-Чепмена
для рассматриваемой сети РО $\{M/M/1$; \, $M/M/1\}$,
при любых исследованных нами комбинациях параметров $\lambda,\, \mu_a,\, \mu_b$
сети
выполняется соотношение: $P_{\text{cond}}<P$.
Это соотношение проиллюстрировано на
рисунке \ref{PictPuass}, где показана зависимость относительной разницы
$\Delta P/P=(P - P_{\text{cond}})/P$
безусловной и условной вероятностей от
$\psi_{a}$ для четырех значений $\psi_{b}$:
0.05;\,
0.35;\,
0.65 и 0.90. Как видно из графиков,
самое значительное уменьшение вероятности происходит в случаях, когда
интенсивности обслуживания в обеих ветвях сети РО совпадают ($\psi_{a}=\psi_{b}$).
В этих случаях доля малых интервалов между событиями (появлениями первых партнеров)
должна уменьшаться, а исходное экспоненциальное распределение -- заметно искажаться,
особенно в окрестности нуля.
Результаты экспериментов по имитационному моделированию при
$\lambda=0.3$ и $\mu_{a}=\mu_{b}=0.8$
(что соответствует значениям $\psi_{a}=\psi_{b}=0.375$)
подтверждают этот вывод.

Для общего случая сети РО $\{M/M/N_a$; \, $M/M/N_b\}$
методами имитационного моделирования построена
таблица \ref{TableIn}, каждая строка которой задает область параметров сети РО,
в которой
входной поток модели $S_0$ является почти пуассоновским (см. Приложение \ref{SectStat})
с точностью $\alpha = 0.01$ и интенсивностью $\lambda_{in} = \lambda$.

\begin{table}[htb]
\caption{
Области почти пуассоновского (с точностью 0.01) потока in$(S_0)$.
}
\begin{center}
\begin{tabular}{|c|c|c|c|}
\hline
$N_a$  &$N_b$   &$\psi_a$   &$\psi_b$\\
\hline
[1, 2]  &[1, 2]  &(0, 0.2] $\cup$ [0.75, 1)    &(0, 0.2]\\
\hline
[1, 2]  &[1, 2]  &(0, 0.2]    &(0, 0.2] $\cup$ [0.75, 1)\\
\hline
[3, 5]  &[3, 5]  &(0, 0.75]    &(0, 1)\\
\hline
[3, 5]  &[3, 5]  &(0, 1)    &(0, 0.75]\\
\hline
[6, $\infty$)  &[6, $\infty$)  &(0, 1)   &(0, 1)\\
\hline
\end{tabular}
\end{center}
\label{TableIn}
\end{table}

\section{Количество заявок в синхронизаторе} \label{SectQuant}

Как видно из таблицы \ref{TableIn}, существуют обширные области параметров
сети РО, в которых входной поток модели $S_0$ синхронизатора можно приближенно
считать пуассоновским.
В этих областях параметров для модели $S_0$ принимаем приближение $S_0 = M/G_0/\infty$.

В этом приближении справедлив следующий точный результат \cite{Adan2002,Kleinrock1979}.
Вероятность того, что в синхронизаторе находится
 $k = 0, ... , \infty$ заявок, ожидающих своих вторых партнеров, равна
\begin{equation}
p_k = (\rho^k/k!) \exp(-\rho),
\label{EqPk}
\end{equation}
где $\rho = \lambda \, \overline{T}$ задает среднее значение количества заявок
в синхронизаторе, а $\overline{T}$ - среднее время пребывания в синхронизаторе.
В работе \cite{SynchroIdeal2007} получены приближенные выражения для функции распределения
вероятности времени пребывания в синхронизаторе и, соответственно, для среднего
времени $\overline{T}$.

Из (\ref{EqPk}), следует, что среднее значение количества заявок, находящихся в
реальном синхронизаторе, ограничено снизу ненулевой величиной $\rho$, которая рассчитана для
идеального (мгновенно срабатывающего) синхронизатора $S_0$. Эта величина характеризует
минимальный размер памяти, необходимый для работы синхронизатора, зависит только
от параметров сети РО, содержащей синхронизатор, и не может быть уменьшена
за счет повышения производительности реального синхронизатора.

\section{Поток на выходе синхронизатора} \label{SectOut}

Поток заявок на выходе модели $S_0$
(то есть виртуальный поток out$(S_0)$)
совпадает с потоком
вторых партнеров маркированных пар.
Как и в случае входного потока in$(S_0)$ в разделе \ref{SectIn},
можно показать, что интенсивность
потока out$(S_0)$ на выходе синхронизатора равна $\lambda_{\text{out}} = \lambda$.

Для общего случая сети РО $\{M/M/N_a$; \, $M/M/N_b\}$
методами имитационного моделирования построена
таблица \ref{TableOut},
каждая строка которой задает область параметров сети РО,
в которой
выходной поток модели $S_0$ является почти пуассоновским (см. Приложение \ref{SectStat})
с точностью $\alpha = 0.01$ и интенсивностью $\lambda_{\text{out}} = \lambda$.

\begin{table}[htb]
\caption{
Области почти пуассоновского (с точностью 0.01) потока out$(S_0)$.
}
\begin{center}
\begin{tabular}{|c|c|c|c|}
\hline
$N_a$  &$N_b$   &$\psi_a$   &$\psi_b$\\\hline
[1, 2]  &[1, 2]  &(0, 0.2] $\cup$ [0.75, 1)    &(0, 0.2], если $|\psi_b - \psi_a|\ge0.1$\\\hline
[1, 2]  &[1, 2]  &(0, 0.2], если $|\psi_a - \psi_b|\ge0.1$    &(0, 0.2] $\cup$ [0.75, 1)\\\hline
[3, 5]  &[3, 5]  &(0, 0.75]    &(0, 1)\\\hline
[3, 5]  &[3, 5]  &(0, 1)    &(0, 0.75]\\\hline
[6, $\infty$)  &[6, $\infty$)  &(0, 1)   &(0, 1)\\\hline
\end{tabular}
\end{center}
\label{TableOut}
\end{table}

\section{Обсуждение результатов}\label{SectDiscus}

В процитированных во введении работах процесс объединения парных заявок в сетях РО не рассматривался.
В основном авторы вычисляли максимум из времен прохождения всех ветвей сети РО,
не интересуясь процессами, происходящими в точке $S$ и после нее.
Предложенная нами модель нового функционального элемента, синхронизатора маркированных
пар, позволила не только оценить ресурсы,
необходимые для бесперебойной работы сети РО, но и определить условия, при которых
синхронизатор (вместе с моделью всей сети РО) может встраиваться в более крупные внешние сети
в качестве марковской подсистемы.

\subparagraph*{Оценка ресурсов,
необходимых для функционирования сети РО без потери информации.}

Модель $S_0$ синхронизатора с бесконечным количеством каналов оказывается пригодной для
прогнозирования реальных ограничений на (конечное) количество каналов $K^{max}$,
необходимых на практике.

Пусть в синхронизаторе допускается потеря доли $\epsilon$ от общего числа заявок,
поступающих на вход синхронизатора.
Найдем размер $K^{max}$ памяти синхронизатора такой, который гарантирует,
что потери заявок не превысят этого порога:
$
P(k > K^{max}) < \epsilon.
$
Отсюда с помощью (\ref{EqPk}) получаем уравнение для нахождения $K^{max}(\epsilon, \rho)$:
$$
1 - \sum_{n=0}^{K^{max}}\frac{\rho^n}{n!}exp(-\rho) < \epsilon.
$$

Аналогично можно найти параметры ветвей  $a$ и $b$, при которых обеспечивается
заданный уровень потерь информации.

\subparagraph*{Оптимизация ресурсов сети с использованием модели синхронизатора.}

Приведем несколько примеров перераспределения ресурсов между составными
частями сети РО: синхронизатором, ветвью $a$ и ветвью $b$.

Предположим, что в ветвях $a$ и $b$,
находится $q_a$ и $q_b$ заявок соответственно,
а в памяти синхронизатора находится $K(S_0)$ заявок.
Пусть размещение заявок в ветвях $a$ и $b$,
а также заявок в памяти синхронизатора $S_0$
реализовано при помощи общей памяти ограниченного объема $M^{max}$:
$$
q_a^{max} + q_b^{max} + K^{max}(S_0) = M^{max}.
$$
В этом случае легко разрешима задача выбора такого набора параметров сети РО
($N_a,\, N_b$), при котором общие потери $\epsilon$ информации в сети принимают минимальное значение.

Если исходить из фиксированного допустимого порога $\epsilon$ потерь информации, можно
найти минимальное значение объема $M^{max}$ общей памяти, при котором обеспечивается
заданный порог потерь.

Если же цена реализации каждого из параллельных каналов обслуживания
существенно различается в ветвях $a$ и $b$, то можно решать задачу оптимизации
цены всей сети (включая синхронизатор)
при заданном значении $\epsilon$ или при заданном значении $M^{max}$.

Теперь рассмотрим случай, когда среднее время $\overline{T}$ пребывания в синхронизаторе
велико по сравнению с $1/\lambda$. Это означает что одна из ветвей (например, $a$)
сети РО работает значительно
быстрее, чем другая. Тогда ветвь $b$ становится узким местом. Для повышения производительности
сети имеет смысл увеличить $N_b$. Если же общая цена важнее производительности, то есть смысл
уменьшить $N_a$. В последнем случае время пребывания в сети РО почти не изменится, уменьшение
параметра $N_a$ приведет лишь к перераспределению общей памяти между средними значениями
количества $q_a$ заявок в ветви $a$ и количества заявок $K(S_0)$ в синхронизаторе.

\subparagraph*{Встраиваемость модели синхронизатора во внешнюю сеть.}

В разделе \ref{SectOut} найдена обширная область параметров сети РО,
в которой выходной поток модели $S_0$ синхронизатора
можно приближенно считать пуассоновским, при условии, что на вход сети поступает
пуассоновский поток. Это означает, что
предложенная модель синхронизатора с параметрами из найденной области
может встраиваться (вместе с моделью всей сети РО) в более крупные внешние сети
в качестве марковской подсистемы. Это свойство марковости существенно
упрощает дальнейший анализ внешней сети.

Нам приятно поблагодарить
А.В. Колодзея,
А.В. Князева,
К.Ю. Платова,
И.А. Кравченко
и А.Ф. Ронжина
за полезные обсуждения.

\appendix

\section{Проверка статистических гипотез}\label{SectStat}

\begin{table}[htb]
\caption{Значения статистических критериев}
\begin{center}
\begin{tabular}{|c|c|}
\hline
&
\begin{tabular}{p{0.5in}|p{0.5in}|p{0.5in}|p{0.5in}|p{0.5in}|p{0.5in}}
&
&
\multicolumn{2}{c|}{in($S_{0}$)} &
\multicolumn{2}{c}{out($S_{0}$)}\\
\hline
$\psi_{a}$&
$\psi_{b}$&
$\chi^{2}$&
$St$&
$\chi^{2}$&
$St$
\end{tabular} \\\hline
\parbox{0.7in}
{
$\lambda=0.3$,  \\
$N_{a}=1$, \\
$N_{b}=1$
}&
\begin{tabular}{p{0.5in}|p{0.5in}|p{0.5in}|p{0.5in}|p{0.5in}|p{0.5in}}
 0.75 &  0.75 &  \textbf{167} & 1.86 &  \textbf{192} & 0.46\\\hline
 0.75 &  0.5 &  \textbf{130} & 0.53 &  \textbf{118} & 2.32\\\hline
 0.75 &  0.25 &  \textbf{61} & 2.02 &  \textbf{56.3} & 2.08\\\hline
 0.75 &  0.2 &  46 & 1.89 &  43.3 & 2.12\\\hline
 0.1 &  0.2 &  35 & 2.09 &  49.2 & 1.83\\\hline
 0.1 &  0.1 &  46.6 & 1.80 &  \textbf{55.4} & 1.98\\\hline
 0.375 &  0.375 &  \textbf{110} & \textbf{2.36} &  \textbf{212.3} & 1.5
 \end{tabular}\\
\hline
\parbox{0.7in}
{
$\lambda=1.5$,  \\
$N_{a}=3$, \\
$N_{b}=5$
}&
\begin{tabular}{p{0.5in}|p{0.5in}|p{0.5in}|p{0.5in}|p{0.5in}|p{0.5in}}
 0.83 &  0.3 &  30 &  0.30 & 37.82 & 0.56\\\hline
 0.91 &  0.6 &  44.2 & 1.57 &  31.33 & 2.20\\\hline
 0.83 &  0.75 &  \textbf{51.9} & \textbf{3.42} &  \textbf{60.02} & \textbf{3.26}\\\hline
 0.83 &  0.83 &  \textbf{54.5} & \textbf{3.04} &  \textbf{74.13} & \textbf{3.53}\\\hline
 0.625 &  0.6 &  45 & 1.38 &  48.47 & 2.17\\\hline
 0.5 &  0.5 &  42.3 & 1.07 &  23.07 & 2.17\\\hline
 0.25 &  0.25 &  30 & 1.68 &  19.17 & 0.45\\\hline
\end{tabular}\\
\hline
\parbox{0.7in}
{
$\lambda=2$,  \\
$N_{a}=8$, \\
$N_{b}=8$
}&
\begin{tabular}{p{0.5in}|p{0.5in}|p{0.5in}|p{0.5in}|p{0.5in}|p{0.5in}}
 0.5 &  0.36 &  33.3 & 0.28 &  33.56 & 0.28\\\hline
 0.83 &  0.625 &  31.75 & 2.09 &  24.07 & 2.07\\\hline
 0.93 &  0.71 &  35.15 & 1.95 &  28.20 & 1.92\\\hline
 0.83 &  0.83 &  25.4 & 1.89 &  45.39 & 2.05\\\hline
 0.9 &  0.93 &  32.7 & 2.20 &  48.26 & 0.65\\\hline
 0.42 &  0.83 &  30.5 & 0.20 &  38.78 & 2.17
 \end{tabular}\\
\hline
\end{tabular}
\end{center}
\label{TableWork}
\end{table}

\textbf{Определение.} При описании результатов имитационного моделирования будем называть
поток заявок \emph{почти пуассоновским с точностью $\alpha$}, если
одновременно принимаются (с уровнем значимости $\alpha$) две следующие статистические гипотезы:
\begin{enumerate}
\item распределение интервалов между моментами поступления заявок является экспоненциальным
(согласно критерию Пирсона);
\item коэффициент корреляции соседних интервалов между моментами поступления заявок равен нулю
(согласно критерию Стьюдента).
\end{enumerate}

Для моделирования случайных процессов в сети РО (рис. \ref{PictForkJoinDouble})
применялся входной поток из $10^5$ заявок с экспоненциальным распределением временных
интервалов, полученный с использованием датчика случайных чисел.
Прямая проверка этого входного потока
показала, что он является почти пуассоновским с точностью $\alpha = 0.01$.
При проверке статистических
гипотез о совпадении эмпирического распределения с гипотетическим
(то есть экспоненциальным) по критерию Пирсона,
область определения плотности распределения разбивалась на 30 отрезков равной вероятности.
Проверялось наличие свойств пуассоновского потока
у эмпирически наблюдаемых
входного in($S_0$) и выходного out($S_0$) потоков заявок в модели синхронизатора $S_0$.
Для этого при многих комбинациях значений параметров сети РО
($\lambda,\, N_a,\, N_b,\, \psi_a,\, \psi_b$) рассчитывались: значение $\chi^2$
критерия Пирсона;
а также значение $St$ критерия
Стьюдента для коэффициента корреляции соседних интервалов в потоке заявок.
Для обоих критериев принимался уровень значимости $\alpha = 0.01$.
При этом граничным значением критерия Пирсона
является величина $\chi_0^2 = 49.6$,
а граничным значением критерия Стьюдента --
величина $St_0 = 2.33$.

В таблице \ref{TableWork} приведены данные, рассчитанные при некоторых значениях
параметров сети. Столбцы, обозначенные
``in($S_0$)'' и ``out($S_0$)'', относятся  к проверке двух гипотез (см. определение
почти пуассоновского потока) о свойствах потоков заявок.
Жирным шрифтом выделены те значения критериев, при которых
соответствующие гипотезы были отвергнуты: $\chi^2 > \chi_0^2$ или $St > St_0$.
Аналогичные данные были использованы для построения таблиц
\ref{TableIn} и \ref{TableOut}.

\end{document}